\documentstyle[11pt,paspconf]{article}

\begin{document}

\title{Peculiar Velocities from Type Ia Supernovae}
\author{Adam G. Riess}
\affil{Department of Astronomy, University of California, Berkeley, CA 
94720-3411}

\begin{abstract}

                Type Ia supernovae have only recently been employed to
                measure the peculiar motions of galaxies. The great distances to which
                SNe Ia can be seen makes them particularly well-suited to constraining
                large-scale velocity flows. The high precision of SN Ia distance
                measurements limits the contamination of such measurements by
                Malmquist or sample selection biases. I review the recent use of
                SNe Ia to measure the motion of the Local Group (updated to include 44
                SNe Ia). The direction of the best-fit motion remains consistent with that
                inferred from the CMB (though modest bulk flows cannot be ruled out) and
 inconsistent with the Lauer-Postman frame.
                Comparisons between SN Ia peculiar velocities and gravity maps from
                IRAS and ORS galaxies yields constraints on the mass density which
                gives rise to the gravitationally induced peculiar motions. Indications are
                for $\beta$=0.4 from IRAS, and 0.3 from the ORS, with $\beta>0.7$
                and $\beta<0.15$ ruled out at 95\% confidence levels from the IRAS
                comparison. The contributions of SNe Ia to peculiar velocity studies are
                limited by statistics, but the great increase in the
                recent discovery rate suggests
                SNe Ia may become increasingly important in constraining flows.
\end{abstract}


\section{What's been done}

    Type Ia
supernova (SNe Ia) are a newcomer to the toolbox used for measuring
peculiar velocities and flows.  Yet, with the rapid increase in the
sample size and precision of distance estimates to these luminous
disruptions of white dwarf stars, SNe Ia show great promise in this
field.  

    I will begin with a review of the published literature (of only
4 or 5 articles) of the
applications of SNe Ia to flow measurements.

    The first attempted use of SNe Ia for flow measurements began in
    the ``low-precision era'' which lasted until the early
    1990s.  This era was characterized by the use of photographic
    photometry and the assumption that SNe Ia were perfect
 ``standard candles'' with homogeneous luminosity and colors.  Such
    data and philosophy yielded distance estimates with approximately
 25\% uncertainty.  Additionally, the sample of SNe Ia were
    concentrated at much closer distances than the present (and
    future) sample.  

    Using 28 primarily photographically observed SNe Ia, Miller \& Branch (1992) were able
    to discern the gravitational influence of the Virgo cluster,
    i.e. Virgocentric infall.  Because the average depth of their sample was
    only about $cz$=2000 km s$^{-1}$, their analysis was insensitive
    to the motion of the Local Group and the influence of the Great
    Attractor.  

    Jerjen \& Tammann (1993) analysed a similar sample of 14 SNe Ia with
    an average depth of $cz$=3000 km s$^{-1}$ (again under the
    assumption of homogeneity of luminosity) but could not detect the
    motion of the Local Group.  

    By 1996, work by Phillips (1993), the Cal\'{a}n/Tololo Search
    (Hamuy et al. 1996a,b,c,d; Maza et al. 1995) the CfA Group
    (Riess, Press, \& Kirshner 1995a, 1996) and others (see Branch
    1998 for a review) demonstrated that with
    high-quality CCD light curves and application of relations between
    the peak luminosity, light curve shapes and color curve shapes, individual
    distance estimates to SNe Ia could reach observed precisions of
    5-7\%.  These improvements and the growing sample of CCD light
    curves ushered in the ``high precision era''.  

    An analysis in my thesis (Riess, Press, \& Kirshner
    1995b) of 13 new SNe Ia from the Calan/Tololo Search made the first
    detection of the motion of the Local Group using SNe Ia.  The
    sample had an effective depth of $cz$=7000 km s$^{-1}$ and a
    typical distance precision of 6\%.  At this time, no corrections
    were applied for host galaxy extinction, though the members of the
    sample exhibited little reddening.  Interestingly, the SN Ia
    measurement was strongly inconsistent with the large bulk flow
    observed from brightest cluster galaxies by Lauer \& Postman (1994), a significant
    result since it was the only
    other sample at a similar depth.  Nearly all of the observed
    disagreement occured in the Galactic $\hat{z}$ direction.  
 Despite the likely effects of
    correlations of small-scale flows (Feldman \& Watkins 1995), the
    measurements remained in conflict.  However, the relative
    imprecision of the SN Ia measurement could not rule out more moderate bulk
    flows on these scales.  

    Recently I have updated this measurement using the light
    and color curves of 44 SNe Ia with effective $cz$=5000 km
    s$^{-1}$.  This sample has been corrected for host galaxy
    extinction using the multicolor light curve shape method (Riess, Press, \& Kirshner 1996).
    The results, show in Figure 1, are highly consistent with the
    previous SN Ia measurement, but have greater precision. The
    best-fit dipole is consistent with the CMB dipole.  Relocating the
    SNe Ia into the CMB frame results in no measurable bulk flow (the
    debiased flow is negligible) with a 1$\sigma$ uncertainty of
    150 km s$^{-1}$.  

     An analysis by Riess, Davis, Baker, \& Kirshner (1997) compared
     the observed peculiar motions of 25 SNe Ia with $cz < 10,000$ km
    s$^{-1}$ to those predicted from the IRAS and ORS gravity maps
     (Nusser \& Davis 1994).
     The predicted peculiar velocities of SNe Ia are a function of the
     local mass in the Universe ($\Omega_M$) as well as the degree to
     which the positions of galaxies indicate the location of mass
     (i.e., the bias parameter).  Together these unknowns
     are quantified by the density parameter,
     $\beta=\Omega_M^{0.6}/b$.  The comparison of the observed and
     predicted peculiar velocities of SNe Ia yields a statistically
     adequate match as well as strong constraints on the value of
     $\beta$.  The fact that the observed and predicted peculiar
     velocity estimates concur (for the best-fit $\beta$) supports the
     gravitational instability paradigm as the source of peculiar
     flows.  The results of the analysis are $\beta=0.40\pm0.15$ from
     the IRAS comparison (and $\beta=0.30\pm0.15$ from the ORS
     comparison, reflecting the relative biasing of infrared and
     optically selected galaxies).  Bootstrap resamplings of the
     gravity maps and the SN Ia sample confirms the validity of the
     uncertainties.  

    Although mentioned previously by others in this conference,
 for completeness we mention an analysis by Zehavi, Riess,
    Kirshner, \& Dekel (1998) which gives a marginal indication of a
    so-called ``Hubble Bubble''.  From 44 SNe Ia, Zehavi et al. (1998)
    found an indication at the 2-3$\sigma$ confidence level of a local
    excess expansion of 6\% within 7000 km s$^{-1}$.  This increase in
    the global Hubble expansion appears to be compensated by a
    small decrease beyond this depth after which the Hubble expansion
    appears to settle to its global value.  The model proposed by
    the authors is that we may live within a local void bounded by a
    wall or density contrast at $\sim$100 Mpc.  More SNe Ia (and other
    distance indicators) will be required to test this provocative
    result.   

    \section{The Future}

    Type Ia supernovae are an attractive tool for contributing to
    the measurement of peculiar velocities and flows in the future.
    SNe Ia provide independent means to measure flows at depths
    unreachable by many other distance indicators.  The individual
    distance precision of SNe Ia results in a reduction of
    systematic errors like Malmquist and sample selection biases which
    often plague peculiar velocity studies.  Individual SNe Ia can be
    corrected for line-of-sight extinction, eliminating a reliance on
    Milky Way extinction maps or inclination corrections.  Finally,
    the pace at which SNe Ia are discovered
    is growing.  By 1999 July 5 SN 1999da had already been discovered,
     starting the 5th cycle through the alphabet after
    only half a year!  In 1998, 20 new SNe Ia with $cz < 0.1$ were added
    to the sample which is useful for peculiar velocity studies.

    A new era of nearby supernova searches is underway.  Below we list
    in ``bullet-form'' searches and collection programs including some of their members,
    facilities, start dates, and successes.

\noindent \hspace*{5mm} $\bullet$ Mount Stromlo \& Siding Springs Observatory SN Search (Schmidt, Germany, Stubbs, Reiss) \\
Up since 1996, 2.3 m at MSSSO, $> 20$ SNe Ia at $z < 0.1$ so far... \\
\hspace*{5mm} $\bullet$ Beijing Astronomical Observatory SN Search
(Li, Qiu, Hu, etc.) \\
Up since 1996, 0.6 m at BAO, 13 SNe Ia at $z < 0.1$ found so far... \\
\hspace*{5mm} $\bullet$ Lick SN Search (Filippenko, Li, Treffers, etc.) \\
Up since 1997, KAIT robotic telescope, 16 SNe Ia at $z < 0.1$ found so far... \\
\hspace*{5mm} $\bullet$ Supernova Cosmology Project Nearby Search
(Perlmutter, Aldering, etc.). \\
Started in 1999, many telescopes, $\sim7$ SNe Ia at $z < 0.1$ found so far... \\
\hspace*{5mm} $\bullet$ CfA Program (Kirshner, Jha, Garnavich, Schmidt, Riess, etc.) \\
Collecting since 1993, $\sim$ 50 SNe Ia collected so far... \\
Others:Perth, EROS, Wise, Tenagra, J. Maza, T. Puckett, W. Johnson, etc \\

   What have the past and present searches produced so far?  I have
   compiled the list of all SNe Ia to date which met the following
   requirements: \\
   $\bullet$CCD photometry \\
   $\bullet$$z < 0.1$ \\
   $\bullet$enough observations recorded to yield precise distances \\

   This list has 115 SNe Ia.  Their positions on the sky and depth can
   be see in Figure 2.  About half of these data have been published
   already (29 from Hamuy et al. 1996; 29 from Riess et al. 1998, 1999;
   15 others in the literature) and the rest are ``in the cans'' of the
   various searches listed above.  The average depth of this sample is
   11,000 km s$^{-1}$ and the effective depth for flow measurements is
   5,000 km s$^{-1}$.  There are 60 objects with $cz < 10,000$ km
   s$^{-1}$.  By looking at Figure 2 we note a few points.  Although
   the distribution between Galactic North and South is not heavily skewed,
   there are more objects in the North.  The typical depth in the
   South is somewhat greater.  The zone of avoidance has been strongly
   avoided to date.  Some concentrations like the Perseus-Pisces
   Supercluster and Coma are not probed while others (Virgo and
   Fornax) are well probed.

   Although there has been little coordination in the past between
   searches, the results are impressive.  This sample, and the ever
   growing future sample, will be a powerful data set with which to
   measure the peculiar motions of test particles subject to gravity in
   the Universe.

\acknowledgments

I wish to express my thanks to Lisa Germany, Greg Aldering, Saurabh Jha, and
Weidong Li for providing lists of discovered SNe Ia.

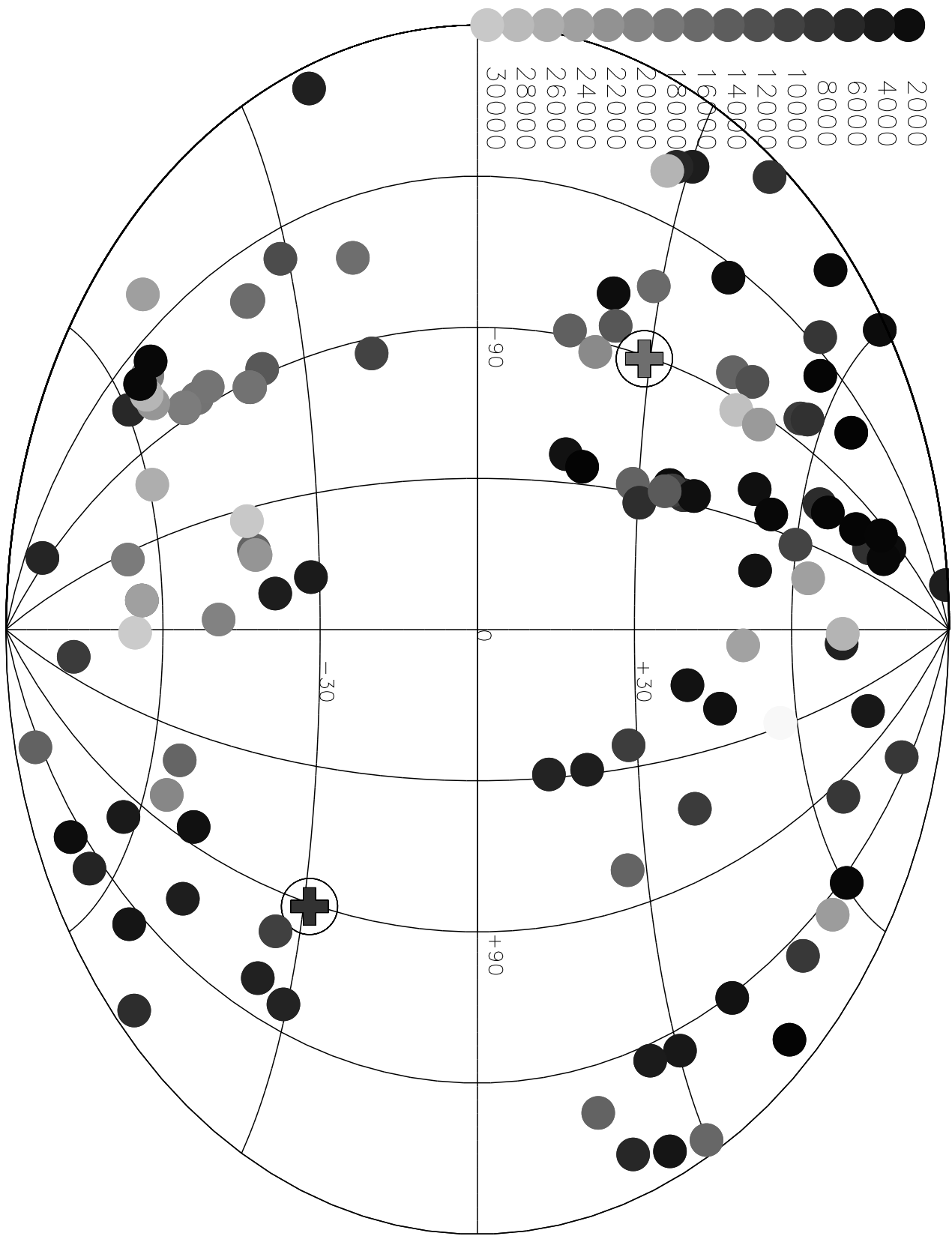
\end{document}